\documentclass[12pt]{iopart}

\usepackage{iopams}  

\begin{document}
\title{Particle interactions predicted from minimum information}

\author{P. A. Mandrin}
\address{Department of Physics, University of Zurich, Winterthurerstrasse 190, 8057 
Z\"urich, CH}
\ead{pierre.mandrin@uzh.ch}

\begin{abstract}
A systematic structure of particle interactions is predicted within and beyond the standard model. The proof is performed either on the basis of (A) a generalisable form of general relativity or, equivalently, (B) minimum information quantum gravity. The emerging structure comprises several chains of interaction generations, one generation being partly realised by the electroweak and the strong interaction. Further interactions have not yet been observed but could be observable in high energy particle collision experiments in the future.
\end{abstract}

\maketitle
%

%%%%%%%%% Introduction
\section{Introduction}
\label{sec:intro}

This article predicts new interactions on the basis of general relativity (GR) and local invariance under phase transformations of the matter fields. Using systematic 3-fold space-time isometries is necessary and sufficient to predict the multiplicity of fields of the standard model (SM) plus higher level interactions and fields. From this point of view, the stringent and exhaustive coverage and extension of the SM is particularly impressive. Such new physics at collision energies below the {\mbox 1 TeV} scale would correspond to the expectations within the perturbative framework \cite {Jaeckel_Khoze}. Alternatively, the proof can be based on minimum information quantum gravity (MIQG) \cite{Mandrin1}\cite{Mandrin2}\cite{Mandrin2a}\cite{Mandrin3}\cite{Mandrin3a}. 

\paragraph{}
MIQG was developed in order to circumvent many problems \cite{Carlip1}\cite{Carlip2} of conventional attempts to quantise gravity and the ambiguity of the quantisation method (as for example precanonical \cite{Kanatchikov1}\cite{Kanatchikov2}\cite{Kanatchikov3} versus canonical quantisation \cite{Thiemann}). MIQG generalises gravitational thermodynamics from \cite{Bekenstein}\cite{Hawking} to ordinary space-time which is assumed to be fundamentally related to statistical mechanics \cite{Hadad_Kupferman}. Unlike other theories of emergent space-time and gravity \cite{Sindoni}, MIQG is based on very few assumptions (quantum statistics, macroscopic space-time and minimum input degrees of freedom).

%%%%%%%%% Gravitational action and symmetries

\section{Gravitational action and symmetries}
\label{sec:gravitation}

Consider a space-time region $\mathcal{M}$ with a neighbourhood in approximate thermal equilibrium \cite{Mandrin3}, boundary $\partial\mathcal{M} = \cup_{k=1}^{k_{\rm max}} \Sigma_k \cup \mathcal{T}$, its space-like part(s) $\Sigma_k$ and time-like part $\mathcal{T}$ being piecewise smooth and every piece normal to each other. Then, the gravitational Einstein-Hilbert action has the boundary variation term \cite{Brown_York_1992}\cite{Brown_York_1993}\cite{York}\cite{Creighton_Mann}

\begin{equation}
\label{eq:var_boundary_grav}
\delta S\bigg|_{\partial\mathcal{M}} = \sum_{\mathcal{A}=\Sigma_k,\mathcal{T}} \int_{\mathcal{A}} {\rm d}^3x \ \sqrt{|\gamma|} \ e^I_i \ \delta \tau^i_{\mathcal{A}I},
\end{equation}

\noindent with indices $I$ (Minkowski) and $i$ (Lorentz) referring to the projections onto the local subspaces associated to $\Sigma_k$ and $\mathcal{T}$, $\gamma_{ij}$ denotes the intrinsic 3d-metrics with determinant $\gamma$ and $\tau^i_{\mathcal{A}I} = (2/\sqrt{|\gamma|})(\delta S / \delta\gamma_{ij})e_{jI}$ \cite{Brown_York_1992}\cite{Brown_York_1993}\cite{Creighton_Mann}\cite{Mandrin2, Mandrin2a}.

\paragraph{}
We can identify several approximate symmetries of $\mathcal{M}$ and evaluate the constants of motion on $\partial\mathcal{M}$, by using the ADM-decomposition \cite{Brown_York_1992}\cite{Brown_York_1993}\cite{Creighton_Mann} of (\ref{eq:var_boundary_grav}) \cite{Mandrin3}:

\begin{equation}
\label{eq:var_boundary_ADM}
\delta S\bigg|_\mathcal{T} = \int_\mathcal{T} {\rm d}^3x [ N \delta ({\sqrt{\sigma} \epsilon}) - N^i \delta (\sqrt{\sigma} j_i) + N \sqrt{\sigma} s^i_I \delta e^I_i],
\end{equation}

\noindent with generalised lapse $N$, shift $N^i$, surface energy density $\epsilon$, surface momentum density $j_i$, and stress vector $s^i_I$. 
We choose $\Sigma$ such that the translation or angular translation or ''Lorentz rapidity'' isometry vectors are contained everywhere within the subspaces associated to $\partial\Sigma$ and $\mathcal{T}$, use ${\rm i}T^{-1} = \oint {\rm i} N {\rm d}t$ (if one coordinate is $t$) and define integral quantities as in \cite{Mandrin1}\cite{Mandrin3}. This yields an expression analogous to the vacuum black hole first law of \cite{Creighton_Mann},

\begin{equation}
\label{eq:first_lawJP}
\delta U \approx T \ \delta S + \omega^m \ \delta\mathcal{J}_m - P^m_\mathcal{M} \ \delta V^\mathcal{M}_m,
\end{equation}

\noindent with constants of motion $\mathcal{J}_m$, coordinate index $m$ and Minkowski index $M$. 

\paragraph{}
In the local Minkowski frame, the cartesian $x^\mu$ translation isometry generates 3+1 conserved momenta $\mathcal{J}_{T\mu}=P_\mu$, we call them the $T$-cluster ($\Sigma$ is a quasi-cube). 

\paragraph{}
Then, we transform the space-like coordinates with indices $i \ne j \ne k$ to cylindric coordinates, $(x^i,x^j) \rightarrow (\rho,\varphi)$. The $\varphi$-isometry provides one conserved angular momentum $\mathcal{J}_{R1}$ ($\Sigma$ is a quasi-cylinder). Transforming to spherical coordinates, $(\rho, x^k) \rightarrow (r,\vartheta)$ provides a second conserved angular momentum $\mathcal{J}_{R2}$ ($\Sigma$ is a quasi-sphere). We transform to 4d-sphere coordinates by transforming to ''boost coordinates'', $(t, r) \rightarrow (z,\chi)$, $t=z\cosh{\chi},r=z\sinh{\chi}$. The $\chi$-isometry (proper Lorentz-invariance) induces a third conserved momentum $\mathcal{J}_{R0}$, where $\tau^i_{\mathcal{T}I}$ is now decomposed with respect to $z$ and $(\chi, \varphi, \vartheta)$ while integrating over the 3d-quasi-hypersphere boundary $\partial \mathcal{M}$. This is the $R$-cluster with $(2+1)$ momenta.

\paragraph{}
We also can start with ''boost coordinates'', $(t, x^i) \rightarrow (u,\chi)$. The conserved momentum is the time evolved centre of mass which can be trivially set to zero using the translation isometry. Repeating $(u, x^j) \rightarrow (v,\psi)$ generates a new $\mathcal{J}_{L1}$, and $(v, x^k) \rightarrow (w,\xi)$ generates another $\mathcal{J}_{L2}$. Before the second ''boost'' step, we can also insert $(x^j,x^k) \rightarrow (\rho,\varphi)$ (cylindric). The subsequent ''boost'' step  $(u, \rho) \rightarrow (\upsilon,\zeta)$ yields one last momentum $\mathcal{J}_{L0}$. Interconverting the first and second ''boost'' steps yields the same $\xi$, $\zeta$ and thus no further momentum can be found. This is the $L$-cluster with $(2+1)$ momenta.

\paragraph{}
Thus, we have a total of three clusters accounting for $(3+1)$ and twice $(2+1)$ conserved momenta $\mathcal{J}_m$. For negligible gravitational field, $P_\mu$ and $\mathcal{J}_{R1,2}$ coincide with the ADM momentum and angular momentum \cite{Ashtekar}\cite{ADM}. (\ref{eq:var_boundary_grav}, \ref{eq:first_lawJP}) also hold for non-vanishing torsion and higher order curvature (GRTH) and agree with the expressions of MIQG \cite{Mandrin1}\cite{Mandrin2}\cite{Mandrin2a}\cite{Mandrin3}\cite{Mandrin3a}.

%%%%%%%%% Extremization under constraints
\section{Extremization under constraints}
\label{sec:level0}

\paragraph{}
As in (\ref{eq:first_lawJP}), $\mathcal{J}_m = \mathcal{J}_T, \mathcal{J}_R, \mathcal{J}_L$ are related to other quantities. For any $\mathcal{J}_m \approx j_m \int_\mathcal{T} {\rm d}^3x \sqrt{\sigma}$, we extremise $S$ under constraints

\begin{equation}
\label{eq:constraint}
\xi_{\mathcal{A}\mathcal{K}}(x^i; j_m) = c_{\mathcal{A}\mathcal{K}}(x^i),
\end{equation}

\noindent $\mathcal{K} = 1, \ldots, \dim(j_m)$, with Lagrange multipliers $\lambda_\mathcal{A}^\mathcal{K}(x^i)$:

\begin{eqnarray}
\label{eq:var_constraint}
\delta S\bigg|_{\partial\mathcal{M}}
= & \sum_\mathcal{A} \int_\mathcal{A} {\rm d}^3x \sqrt{|\gamma|} [ e^I_i \delta \tau^i_{\mathcal{A}I} + \delta(\lambda_\mathcal{A}^\mathcal{K} \xi_{\mathcal{A}\mathcal{K}})] & = 0, \\
\label{eq:var_lambda}
& \sum_\mathcal{A} \int_{\mathcal{A}} {\rm d}^3x \ \sqrt{|\gamma|} \ \xi_{\mathcal{A}\mathcal{K}} \ \delta \lambda_\mathcal{A}^\mathcal{K} & = 0.
\end{eqnarray}

\noindent We can obtain a quadratic form by decomposing the expression $\lambda_\mathcal{A}^\mathcal{K} \ \xi_{\mathcal{A}\mathcal{K}}$ as follows (suppressing the label $\mathcal{A}$):

\begin{equation}
\label{eq:quadratic_pi}
\lambda^\mathcal{K} \ \xi_\mathcal{K} = \psi^\dagger_\mathcal{K} \ \alpha^\mu \ \psi^\mathcal{K} \ n_\mu = \pi_\mathcal{K}^\mu \ \psi^\mathcal{K} \ n_\mu = \pi_\mathcal{K} \ \psi^\mathcal{K}.
\end{equation}

\noindent The dagger in $\psi^\dagger_\mathcal{K}$ reflects the undetermined format of $\psi^\mathcal{K}$ and $\alpha^\mu$. As in \cite{Mandrin1}\cite{Mandrin2}\cite{Mandrin3}, $\pi_\mathcal{K}$ is expressed by its flow $\pi_\mathcal{K}^\mu$ across $\mathcal{A}$, while $\mathcal{A}$ has an associated subspace with unit normal vector $n^\mu$, $\pi_\mathcal{K}^\mu = \psi^{\dagger}_\mathcal{K} \ \alpha^\mu$, $\pi_\mathcal{K}^\mu \ n_\mu = \pi_\mathcal{K}$. Insert (\ref{eq:quadratic_pi}) into (\ref{eq:var_constraint}), apply Gauss' theorem, (\ref{eq:var_lambda}), the procedure in \cite{Mandrin3} and obtain the (Legendre transformed) $\mathcal{J}_m$-contribution $\delta S_m$:

\begin{eqnarray}
\label{eq:S_part_pi}
\delta S_m\big|_{\partial\mathcal{M}} & = & \int_{\mathcal{M}} {\rm d}^4x \sqrt{-g} \ [\delta{j'}_\mathcal{K} \psi^\mathcal{K} + ( \delta\pi_\mathcal{K}^\mu) \partial_\mu \psi^\mathcal{K}],  \\
\label{eq:S_particles}
S_m & = & \int_{\mathcal{M}} {\rm d}^4x \sqrt{-g} \ [{j'}_\mathcal{K} \psi^\mathcal{K} + \psi^\dagger_\mathcal{K} \alpha^\mu \partial_\mu \psi^\mathcal{K}].
\end{eqnarray}

\noindent with $\delta{j'}_\mathcal{K} = \nabla_\mu \delta\pi_\mathcal{K}^\mu$. To linear order, the local Lorentzian structure makes $\psi^\mathcal{K}$ oscillatory, suggesting complex notation. 

\paragraph{}
First consider the $\mathcal{J}_{R1,2}$-sector. Imposing invariance on $S$ under local phase transformations $\psi^\mathcal{K} \rightarrow U^\mathcal{K}_\mathcal{L}\psi^\mathcal{L} = {\rm e}^{{\rm i} \omega_I(x){\tau^I}^\mathcal{K}_\mathcal{L}} \psi^\mathcal{L}$ ($\tau^I$ spans the 3-dimensional Clifford algebra) implies the $SU(2)$ invariance of $S$. Proof of the $SU(2)$ invariance from MIQG: $\psi^\mathcal{K}$ changes the number of quanta in $\mathcal{M}$. By construction, the choice of the $\mathcal{K}$-component is arbitrary. Thus, $S$ is invariant under $\mathcal{K}$-rotations of $\psi^\mathcal{K}$ and thus $SU(2)$ invariant.

\paragraph{}
The $\mathcal{J}_{R0}$-sector provides one more constraint with one single field $\psi^0$ and induces the invariance of $S$ under transformations $\psi^\mathcal{K} \rightarrow U^{\mathcal{K}0}\psi^\mathcal{K} = {\rm e}^{{\rm i} \beta(x) Y^{\mathcal{K}}} \psi^\mathcal{K}$, to be imposed on the $\psi^\mathcal{K}$ of the $\mathcal{J}_{R1,2}$- and $\mathcal{J}_{R0}$-sector. Thus, the $R$-cluster yields $U(1) \times SU(2)$ invariance. 

\paragraph{}
Using the same procedure, the $T$-cluster yields $U(1) \times SU(3)$-invariance, and the $L$-cluster yields $U(1) \times SU(2)$-invariance. Consider the $\mathcal{J}_{Ti}$-sector ($i = 1 \ldots 3$). Given a point $p \in \mathcal{T}$, we can choose the cartesian coordinates at $p$ and the value $i$ so that $j_{R1} = \rho \ j_{Ti}$. This gives us three constraints $\xi_{\mathcal{A}\mathcal{K}_T1}(x^i; j_T; j_R)$ and three fields $\psi^{\mathcal{K}_T1}$. We can repeat this step for the second $R$-component, $j_{R2} = r \ j_{Ti}$, to obtain three more constraints and thus three fields $\psi^{\mathcal{K}_T2}$. Extending the procedure to include the $j_{T0}$, $j_{R0}$ and including the $L$-cluster as well leads to a total of $(3+1)\times(1+2)\times(1+2)=36$ fermion fields $\psi^{\mathcal{K}_T\mathcal{K}_R\mathcal{K}_L}$. Altogether, these fields are subject to the symmetry $U(1) \times SU(3) \times U(1) \times SU(2) \times U(1) \times SU(2)$.

\paragraph{}
We now show that the $\mathcal{J}_m$-sectors of each cluster describe spin 1/2 fields.
In a local Minkowski frame, consider the plane wave ansatz $\psi_\mathcal{K} = \psi_{\mathcal{K}0} \ {\rm e}^{{\rm i}p_\nu x^\nu}$, where $p_\mu$ does not depend on $x^\mu$. Insert this ansatz and the factor $\delta (\psi_\mathcal{K}^\dagger \ \zeta) = \nabla_\mu \delta \pi_\mathcal{K}^\mu$ into (\ref{eq:S_part_pi}) ($\hbar = 1$):

\begin{equation}
\label{eq:dS_constr}
\delta S_m\big|_{\partial\mathcal{M}} = \int_{\mathcal{M}} {\rm d}^4x \ [\delta(\psi_\mathcal{K}^\dagger \ \zeta) \ \psi^\mathcal{K} + ( \delta\pi_\mathcal{K}^\mu) \ {\rm i} p_\mu \psi^\mathcal{K}].
\end{equation}

\noindent After replacing ${\rm i}p_\mu \rightarrow \partial_\mu$ and computing $S_m$ \cite{Mandrin3}, $\delta S = 0$ at fixed $\tau^i_{\mathcal{A}I}$ yields

\begin{equation}
\label{eq:ELE}
\psi_\mathcal{K}^\dagger \ \zeta' - (\partial_\mu \psi^\dagger_\mathcal{K}) \ \alpha^\mu = \psi_\mathcal{K}^\dagger \ \zeta' + {\rm i}p_\mu \psi^\dagger_\mathcal{K} \ \alpha^\mu = 0
\end{equation}

\noindent with $\zeta' = \zeta - \nabla_\mu \alpha^\mu$. Solutions of $\delta S = 0$ also solve

\begin{equation}
\label{eq:ELEpsquare}
\psi^\dagger_\mathcal{K} \ (\zeta'  + {\rm i} p_\mu \ \alpha^\mu)(\zeta'  + {\rm i} p^\nu \ \alpha_\nu) = 0.
\end{equation}

\noindent The term quadratic in $p_\mu$ yields $\psi^\dagger_\mathcal{K} \ \alpha^\mu \ \alpha_\nu \ p_\mu p^\nu$. We still are free to choose $\alpha^\mu$ so that $\alpha^\mu \alpha_\mu = 2$. Imposing $p_\mu p^\mu = C$ leads to the anti-commutator relation $\{\alpha^\mu, \alpha_\nu \} = 2 \ \delta^\mu_\nu$, as for the Dirac equation. $\psi^\mathcal{K}$ must therefore be spin 1/2 fields. The solutions also imply $C = (m^\mathcal{K})^2$ for a field of particle mass $m^\mathcal{K}$ in the MIQG interpretation, and we can expand the general field $\psi_\mathcal{K} = \int {\rm d}^4p \ \psi_{\mathcal{K}}(p_\mu) \ {\rm e}^{{\rm i}p_\nu x^\nu}$ accordingly.

% Higher level contstraints
\section{Higher level contstraints}
\label{sec:higherlevel}

The transformation $U^\mathcal{K}_\mathcal{L} = {\rm e}^{{\rm i} \omega_k(x){\tau^k}^\mathcal{K}_\mathcal{L}}$ for the $\psi^\mathcal{K}$ of each $\mathcal{J}_m$-sector (with symmetry group generators $\tau^k$) yields a Noether-current density $j_{(1)}^{k\mu} = [\delta\mathcal{L}/\delta(\partial_\mu \psi^\mathcal{K})]$$\cdot(\delta \psi^\mathcal{K} / \delta \omega_k) - j_{(1)0}^{k\mu}$. We call this the level 1 Noether current density and write the level (1) in parentheses in order to distinguish level 1 quantities from the corresponding above quantities (level 0). Again, $j_{(1)}^{k\mu}$ depends on other quantities, yielding additional constraint equations

\begin{equation}
\label{eq:constraint1}
\xi_{(1)\mathcal{K}_{(1)}i}(x^j; j_{(0)m}; j_{(1)}^{k\mu}) = c_{(1)\mathcal{K}_{(1)}i}(x^j; j_{(0)m}),
\end{equation}

\noindent where $\mathcal{K}_{(1)}$ and $k$ have the same number $n_k$ of values, and writing 

\begin{equation}
\label{eq:indexing}
\xi_{(1)\mathcal{K}_{(1)}i} = A_{\mathcal{K}_{(1)}k}(x^j; j_{(0)m}) \hat{\gamma}^\mu_i \gamma_{\mu\nu} j_{(1)}^{k\nu} + B_{\mathcal{K}_{(1)}i}(x^j; j_{(0)m})
\end{equation}

\noindent (with projector $\hat{\gamma}^\mu_i$) preserves the index $i$ associated to the boundary. Introducing level 1 Lagrange multipliers $\lambda_{(1)}^{\mathcal{K}_{(1)}i}(x^j)$ and using the same procedure as in the last section, we obtain:

\begin{equation}
\label{eq:var_matter_level1}
\delta S\bigg|_{\mathcal{A}} =  \int_\mathcal{A} {\rm d}^3x \ \sqrt{-\gamma} \ [\tau^i_I \ \delta e^I_i + \delta \pi_{(0)\mathcal{K}_T\mathcal{K}_R\mathcal{K}_L} \ \psi_{(0)}^{\mathcal{K}_T\mathcal{K}_R\mathcal{K}_L} + \delta \pi_{(1)C\mathcal{K}_{(1)}i} \ \psi_{(1)}^{C\mathcal{K}_{(1)}i}],
\end{equation}

\noindent where the sum over double cluster indices $C=T,R,L$ is understood in the last term.

\paragraph{}
Again, $S$ must be phase invariant. The transformations involving the index $i$ or $\mu$ are part of the space-time diffeomorphisms and already have induced the $\mathcal{J}_m$. There remains $\psi_{(1)}^{\mathcal{K}'\mu} \rightarrow \hat{U}^{\mathcal{K}'}_{\mathcal{L}'} \psi_{(1)}^{\mathcal{L}'\mu} = {\rm e}^{{\rm i} \vartheta_q(x){\lambda^q}^{\mathcal{K}'}_{\mathcal{L}'}} \psi_{(1)}^{\mathcal{L}'\mu}$, where $\lambda^q$ are either the generators of $SU(n_k)$ or $U(1)$ (if we set $\lambda^{\mathcal{K}'}_{\mathcal{L}'} = \beta_{\mathcal{L}'}\delta^{\mathcal{K}'}_{\mathcal{L}'}$), $\mathcal{K}' = C\mathcal{K}_{(1)}$ and $C$ is any cluster.

\paragraph{}
We can restrict the spin value of $\psi_{(1)}^{C\mathcal{K}\mu}$. To account for the flow of the field across the boundary, we again set $\delta \pi_{\mathcal{K}\mu} = \delta \pi_{\mathcal{K}\mu\nu} n^\nu$ (level and cluster index suppressed). We write $\delta \pi_{\mathcal{K}\mu\nu}$ as a normal to the space associated to $\mathcal{A}$, using a 4d-vector product (asymmetric in $\mu$, $\nu$). With $\omega_\nu = \delta\pi_{\mathcal{K}\mu\nu} \ \psi^{\mathcal{K}\mu}$, Gauss' theorem and $\delta \pi_{\mathcal{K}\mu\nu} = - \delta \pi_{\mathcal{K}\nu\mu}$, we have

\begin{equation}
\label{eq:stokes}
\int_{\partial\mathcal{M}} \omega = \int_{\mathcal{M}} {\rm d}\omega = \int_{\mathcal{M}} {\rm d}^4x \ [(\nabla^{\nu}\pi_{\mathcal{K}\mu\nu}) \ \psi^{\mathcal{K}\mu} + \frac{1}{2}\pi_{\mathcal{K}\mu\nu} \ F^{\mathcal{K}\mu\nu}]
\end{equation}

\noindent with $F^{\mathcal{K}\mu\nu} = \nabla^{\nu} \ \psi^{\mathcal{K}\mu}-\nabla^{\mu} \ \psi^{\mathcal{K}\nu}$. The elements of $F^{\mathcal{K}\mu\nu}$ have the same structure as the electromagnetic field tensor and thus represent a field of integer spin. 

\paragraph{}
The same procedure as for level 1 leads to the level 2 Noether currents and constraints, and so on. This yields:

\begin{equation}
\label{eq:var_matter_all}
\delta S\bigg|_{\mathcal{A}} =  \int_\mathcal{A} {\rm d}^3x \sqrt{|\gamma|} (\tau^i_I \delta e^I_i + \sum_{l\ge 0} \delta \pi_{(l)\{\mathcal{K}_T\mathcal{K}_R\mathcal{K}_L\}[C\mathcal{K}_{(l)}i]}  \ \psi_{(l)}^{\{\mathcal{K}_T\mathcal{K}_R\mathcal{K}_L\}[C\mathcal{K}_{(l)}i]} ),
\end{equation}

\noindent where $[x]$ means $x$ for $l>0$ (else void), and $\{x\}$ means $x$ for $l=0$ (else void). Finally, the total function $S_{\rm total}$ is obtained with \cite{Mandrin2}\cite{Mandrin2a}:

\begin{eqnarray}
S_{\rm total} & = & \int_{\mathcal{M}}  {\rm d}^4x \ \sqrt{-g} \ (e^I_\mu e^J_\nu \Phi^{\mu\nu}_{IJ} + \omega_{\mu IJ} \Omega^{\mu IJ} \nonumber \\
& + & \sum_{l\ge 0} \ {j'}_{(l)\{\mathcal{K}_T\mathcal{K}_R\mathcal{K}_L\}[C\mathcal{K}_{(l)}\mu]} \ \psi_{(l)}^{\{\mathcal{K}_T\mathcal{K}_R\mathcal{K}_L\}[C\mathcal{K}_{(l)}\mu]} \nonumber \\
\label{eq:action_final}
& + & [\frac{1}{2}] \pi_{(l)\{\mathcal{K}_T\mathcal{K}_R\mathcal{K}_L\}[C\mathcal{K}_{(l)}\mu]\nu} \ F_{(l)}^{\{\mathcal{K}_T\mathcal{K}_R\mathcal{K}_L\}[C\mathcal{K}_{(l)}\mu]\nu}),
\end{eqnarray}

\noindent where $\omega_{\mu IJ}$ is the connection 1-form, $\Phi^{\mu\nu}_{IJ}$ is the generalised curvature 2-form, $\Omega^{\mu IJ}$ is the gravitational variable conjugate to the connection 1-form, $F_{(l)}^{\{\mathcal{K}_T\mathcal{K}_R\mathcal{K}_L\}[C\mathcal{K}_{(l)}\mu]\nu}$ is the covariant derivative (antisymmetrised if $l>0$) of $\psi_{(l)}^{\{\mathcal{K}_T\mathcal{K}_R\mathcal{K}_L\}[C\mathcal{K}_{(l)}\mu]}$, ${j'}_{(l)\{\mathcal{K}_T\mathcal{K}_R\mathcal{K}_L\}[C\mathcal{K}_{(l)}\mu]}$ is the generalised current density (the divergence of $\pi_{(l)\{\mathcal{K}_T\mathcal{K}_R\mathcal{K}_L\}[C\mathcal{K}_{(l)}\mu]\nu}$).

\paragraph{}
Increasing levels involve more and more Noether currents entering the ''first law'' (\ref{eq:first_lawJP}), and the energy associated to the fields may increase accordingly. Therefore, the expansion of levels should not have infinitely many levels, but stop before the Planck energy is reached.

%%%%%%%%%%%%% Identification and prediction of interactions

\section{Identification and prediction of interactions}
\label{sec:predict}

\subsection{Level 0}

The $(3+1)\times(1+2)\times(1+2)=36$ spin 1/2 fields $\psi_{(0)}^{\mathcal{K}_T\mathcal{K}_R\mathcal{K}_L}$ have the same rank as the formal Dirac functions of the standard model: (3 quark colours + 1 lepton)$\times$(1 + 2 electroweak components: singlet plus doublet)$\times$(3 flavours). The spinor structure arises once $\delta S = 0$ is solved. The same $U(1) \times SU(2)$ and $SU(3)$ invariances are found as for the electroweak and strong interaction mechanisms. These coincidences indicate that the level 0 might have to be identified with the most elementary known fermions, and one additional structure ($U(1) \times SU(2)$) might explain the flavours.

\subsection{Level 1}

The boson fields $\psi_{(1)}^{C\mathcal{K}_{(1)}\mu}$ have the same rank and index structure as the set of electroweak interaction potentials $(B^\mu, W_K^\mu)$ associated to the $\gamma$, the $Z^0$, $W^+$ and $W^-$ bosons ($U(1)\times SU(2)$), the gluons ($SU(3)$), one more singlet coupling to leptons and quarks (as does the Higgs-field), and a third family of 4 bosons ($U(1)\times SU(2)$) (flavour?). To lowest order, the level 1 bosons are exchanged between level 0 fermions, and non-linearities may allow for couplings between the bosons. We find a good correspondence with the standard model.

\subsection{Higher level interactions}

All other interactions predicted according to our expansion (\ref{eq:action_final}) have not yet been observed. The particle collision energy might be required to be higher than for the level 0 and 1 processes in order for the higher level particle resonances to be observed.
For example, consider $\psi_{(2)}^{C\mathcal{K}_{(2)}\mu}$ from the $SU(3)$ symmetry of the electroweak sector $SU(2)$. These fields represent 8 bosons $X^{\mathcal{K}_{(2)}}$ exchanged between the bosons $W^*$ ($\gamma, Z^0, W^+, W^-$): $W^* \rightarrow {W^*}' + X^{\mathcal{K}_{(2)}}$.

%%%%%%%%%%%%% 5. Conclusions

\section{Conclusions}
\label{sec:conclusion}

Starting either from well-established general relativity (with relaxed prescriptions on the connections and on the expansion of the differential orders) or from MIQG (defined by three assumptions), it has been shown that several chains of interactions (with new particles) follow. These chains comprise all fields and interactions of the standard model as well as additional structures of interactions. The proof is even more elegant if based on MIQG rather than GRTH, as the assumption of phase invariance may be omitted. It is hoped that the existence of some higher level interactions might be confirmed in high energy collision experiments, via new particle resonances and the associated decay channels.

%%%%%%%%%%%%%%%% Acknowledgements

\section*{Acknowledgements}

I would like to thank Philippe Jetzer for hospitality at University of Zurich.

% Bibloography

%%%%%%%%%%%  References


\section*{References}

\begin{thebibliography}{99}

\bibitem{Jaeckel_Khoze} J. Jaeckel, V. Khoze, arxiv:1411.5633.
\bibitem{Mandrin1} P. A. Mandrin, Int. J. Theor. Phys. \textbf{53} (2014) 4250--4266, DOI 10.1007/s10773-014-2176-8.
\bibitem{Mandrin2} P. A. Mandrin, arxiv:1408.1896 (2014).
\bibitem{Mandrin2a} P. A. Mandrin, Spin-compatible construction of a consistent quantum gravity model from minimum information, Poster presented at the Conference on Quantum Gravity, "Frontiers of Fundamental Physics 2014", Marseille, 15-18 July (2014).
\bibitem{Mandrin3} P. A. Mandrin, arxiv:1409.0809 (2014).
\bibitem{Mandrin3a} P. A. Mandrin, A state occupation number prescription in the scope of minimum information quantum gravity, Talk presented at the conference Conceptual and Technical Challenges for Quantum Gravity 2014, Rome, 8-12 September (2014).
\bibitem{Carlip1} S. Carlip, Rep. Prog. Phys. \textbf{64} (2001) 885, arxiv:0108040 (2001).
\bibitem{Carlip2} S. Carlip, Class. Quantum Grav. \textbf{25} (2008) 154010, arXiv:0803.3456.
\bibitem{Kanatchikov1} I. V. Kanatchikov, J. Phys. Conference Series \textbf{442} (2013) 012041.
\bibitem{Kanatchikov2} I. V. Kanatchikov, Phys. Lett. \textbf{A283} (2001) 25--36. 
\bibitem{Kanatchikov3} I. V. Kanatchikov, Int. J. Theor. Phys. \textbf{40} (2001) 1121--1149.
\bibitem{Thiemann} T. Thiemann, \textit{Modern Canonical Quantum General Relativity}, Cambridge Monographs on Mathematical Physics, Cambridge, 2007.
\bibitem{Bekenstein} J. D. Bekenstein, Phys. Rev. D \textbf{7}  (1973) 2333--2346.
\bibitem{Hawking} S. W. Hawking, Commun. math. Phys. \textbf{43} (1975) 199--220.
\bibitem{Hadad_Kupferman} M. Hadad, J. Kupferman, arxiv:1006.3161 (2014).
\bibitem{Sindoni} L. Sindoni, arxiv.org:1110.0686 (2012).
\bibitem{Ashtekar} A. Ashtekar, Asymptotic Structure of the Gravitational Field at Spatial Infinity, in General Realtivity and Gravitation, Vol. 2, ed. A. Held, Plenum, New York (1980).
\bibitem{ADM} R. Arnowitt, S. Deser, C. W. Misner, The Dynamics of General Relativity, in Gravitation: An Introduction to Current Research, ed. L. Witten, Wiley, New York (1962).
\bibitem{Brown_York_1992} J. D. Brown, J. W. York, arXiv:9209014 (1992).
\bibitem{Brown_York_1993} J. D. Brown, J. W. York, Phys. Rev. D \textbf{47} (1993) 1407--1419.
\bibitem{York} J. W. York, Found. Phys. \textbf{16} (1986) 249--257.
\bibitem{Creighton_Mann} D. E. Creighton, R. B. Mann, Phys. Rev. D \textbf{52} (1995) 4569--4587.

\end{thebibliography}
\end{document}